\input harvmac
\input epsf
%
\def\journal#1&#2(#3){\unskip, \sl #1\ \bf #2 \rm(19#3) }
\def\andjournal#1&#2(#3){\sl #1~\bf #2 \rm (19#3) }

\def\ie{{\it i.e.}}
\def\eg{{\it e.g.}}
\def\cf{{\it c.f.}}

\def\frac#1#2{{#1\over#2}}

\def\half{\frac12}
\def\hf{{\textstyle\half}}

\def\inbar{\,\vrule height1.5ex width.4pt depth0pt}
\def\IC{\relax\hbox{$\inbar\kern-.3em{\rm C}$}}
\def\IR{\relax{\rm I\kern-.18em R}}
\def\IP{\relax{\rm I\kern-.18em P}}

%
%
\def\np#1#2#3{Nucl.\ Phys.\ {\bf B#1} (#2) #3}
\def\pl#1#2#3{Phys.\ Lett.\ {\bf #1B} (#2) #3}

\def\prl#1#2#3{Phys.\ Rev.\ Lett.\ {\bf #1} (#2) #3}

\def\prd#1#2#3{Phys.\ Rev.\ {\bf D#1} (#2) #3}

\def\cmp#1#2#3{Comm.\ Math.\ Phys.\ {\bf #1} (#2) #3}

\def\jhep#1#2#3{J.\ High.\ Energy Phys.\ {\bf #1} (#2) #3}

\catcode`\@=11
\def\slash#1{\mathord{\mathpalette\c@ncel{#1}}}
\overfullrule=0pt

\def\OO{{\cal O}}

\def\a{\alpha}

\def\underrel#1\over#2{\mathrel{\mathop{\kern\z@#1}\limits_{#2}}}

\catcode`\@=12


%

\def\det{{\rm det}}

\def\det{{\rm det}}


\def\[{\hbox{{\titlefont [}}}
\def\]{\hbox{{\titlefont ]}}}

\def\hk{hyperk\"ahler }

\def\nfr#1#2{{\textstyle{\frac{#1}{#2}}}}

\nref\malda{J.~Maldacena, ``The large N limit of Superconformal field
theories and Supergravity'', Adv. Theor. Math. Phys. {\bf 2} (1998)
231, hep-th/9711200.}

\nref\kasi{S.~Kachru and E.~Silverstein, "4-D conformal theories and
strings on orbifolds" \prl{80}{1998}{4855}, hep-th/9802183.}

\nref\lnv{A.~Lawrence, N.~Nekrasov and C.~Vafa, "On conformal field
theories in four dimensions", \np{533}{1998}{199}, hep-th/9803015.}

\nref\ansar{A.~Fayyazuddin and M.~Spalinski, ``Large N Superconformal
Gauge Theories and Supergravity Orientifolds'', \np{535}{1998}{219},
hep-th/9905096.}

\nref\igored{I.~Klebanov and E.~Witten, ``Superconformal field theory
on three-branes at a Calabi-Yau singularity'', \np{536}{1998}{199},
hep-th/9807080.}

\nref\dm{M.~R.~Douglas and G.~Moore, ``D-branes, Quivers, and ALE
Instantons'', hep-th/9603167.}

\nref\jm{C. V. Johnson and R. C. Myers, ``Aspects of Type
IIB Theory on ALE Spaces'',
\prd{55}{1997}{6382}, hep-th/9610140.}

\nref\gr{M.~R.~Douglas, B.~R.~Greene and D.~Morrison, ``Orbifold
resolution by D-Branes'', \np{509}{1997}{84}, hep-th/9704151.}

\nref\bdual{A.~Hanany and A.~Uranga, ``Brane Boxes and Branes on
Singularities'', \jhep{05}{1998}{013}, hep-th/9805139.}

\nref\ogivaf{H.~Ooguri and C.~Vafa, ``Geometry of $N=1$ Dualities in
Four Dimensions'', \np{500}{1997}{62}, hep-th/9702180.}

\nref\hw{A.~Hanany and E.~Witten, ``Type IIB Superstings, BPS
monopoles, and three-dimensional gauge dynamics'',
\np{492}{1997}{152}, hep-th/9611230.}

\nref\adual{A.~Karch, D.~L\"{u}st and D.~Smith, ``Equivalence of
geometric engineering and Hanany-Witten via fractional branes'',
\np{533}{1998}{348}, hep-th/9803232.}

\nref\ddual{A.~Kapustin, ``$D_{N}$ quivers from branes'',
\jhep{12}{1998}{015}, hep-th/9806238.}

\nref\dddual{A.~Hanany and A.~Zaffaroni, ``Issues on Orientifolds: On
the Brane Construction of Gauge Theories with $SO(2N)$ Global
Symmetry'', \jhep{07}{1999}{009}, hep-th/9903242.}

\nref\lr{U. Lindstr\"om and M. Ro\v cek, ``Scalar Tensor
Duality and $N=1,2$ Non-linear $\sigma$-models'',
\np{222}{1983}{285}.}

\nref\hklr{N. J. Hitchin, A. Karlhede, U.\ Lindstr\"om, and
M.\ Ro\v cek, ``Hyperk\"ahler Metrics and Supersymmetry'',
\cmp{108}{1987}{535}.}

\nref\nh{N. J. Hitchin, ``Polygons and gravitons'', Math. Proc. Camb.
Phil. Soc. {\bf 85} (1979) 465.}

\nref\pk{P. B. Kronheimer, ``The Construction of ALE spaces
as Hyper-K\"ahler Quotients'', J. Diff. Geom. {\bf 29}
(1989) 665.}

\nref\abfgz{D. Anselmi, M. Bill\'o, P. Fr\'e, L. Girardello,
and A. Zaffaroni, ``ALE Manifolds and Conformal Field
Theories'', Int. J. Mod. Phys. {\bf A9} (1994) 3007,
hep-th/9304135}

\nref\gns{Steven Gubser, Nikita Nekrasov, and Samson Shatashvili,
``Generalized conifolds and 4-Dimensional N=1 SuperConformal Field
Theory'', \jhep{05}{1999}{003}, hep-th/9811230.}

\nref\chklr{C. M. Hull, A. Karlhede, U. Lindstr\"om, and M. Ro\v cek,
``Nonlinear $\sigma$-models and their gauging in and out of
superspace'', \np{266}{1986}{1}.}

\nref\bz{B. Zumino, ``Supersymmetry and K\"ahler manifolds'',
\pl{87}{1979}{203}.}


\nref\ns{N.~Seiberg, ``Electric-Magnetic Duality in Supersymmetric
Non-Abelian Gauge Theories'', \np{435}{1995}{129}, hep-th/9411149.}

\nref\mn{J.~A.~Minahan and D.~Nemeschansky,
``An N=2 superconformal fixed point with E(6) global symmetry'',
\np{482}{1996}{142}, hep-th/9608047.}

\nref\mnII{J.~A.~Minahan and D.~Nemeschansky,``Superconformal
fixed points with E(n) global symmetry'',
\np{489}{1997}{24}, hep-th/9610076.}

\nref\nty{M.~Noguchi, S.~Terashima, and S-K Yang, "N=2 Superconformal
Field Theory with ADE Global Symmetry on a D3-brane Probe",
hep-th/9903215.}

\nref\ec{E. Calabi, ``Metriques K\"ahl\'eriennes et fibr\'e
holomorphe", Ann. Sci. Ec. Norm. Super. {\bf12} (1979) 269.}

\nref\mp{A. J. Macfarlane and H. Pfeiffer, ``On characteristic equations, trace identities
and Casimir operators of simple Lie algebras'', math-ph/9907024.}

\rightline{USITP-99-07}
\vskip-.1cm
\rightline{YITP-99-46}
\vskip-.1cm
\Title{\rightline{hep-th/9908082}}
{\vbox{\centerline{Hyperk\"ahler quotients and algebraic curves}}}
\vskip-.5cm
\centerline{Ulf Lindstr\"om\footnote{$^*$}{ul@physto.se}}
\centerline{\it Institute of Theoretical Physics,
University of Stockholm}
\vskip-.1cm
\centerline{\it Box 6730}
\vskip-.1cm
\centerline{\it S-113 85 Stockholm, SWEDEN}
\medskip
\centerline {Martin Ro\v cek\footnote{$^{**}$}
{rocek@insti.physics.sunysb.edu}}
\centerline{\it C.N. Yang Institute of Theoretical Physics,
State University of New York }
\vskip-.1cm
\centerline{\it Stony Brook, NY 11794-3840, USA }
\medskip
\centerline{Rikard von Unge\footnote{$^{\dagger}$}
{unge@physics.muni.cz}}
\centerline{\it Department of Theoretical Physics and Astrophysics}
\vskip-.1cm
\centerline{\it Faculty of Science, Masaryk University}
\vskip-.1cm
\centerline{\it Kotl\'{a}\v{r}sk\'{a} 2, CZ-611 37, Brno, Czech Republic}
\bigskip
\noindent
Abstract:
{We develop a graphical representation of polynomial invariants of
unitary gauge groups, and use it to find the algebraic curve
corresponding to a \hk quotient of a linear space. We apply this method
to four dimensional ALE spaces, and for the $A_k$, $D_k$, and $E_6$
cases, derive the explicit relation between the deformations of the
curves away from the orbifold limit and the Fayet-Iliopoulos parameters
in the corresponding quotient construction. We work out the
orbifold limit of $E_7$, $E_8$, and some higher dimensional examples.}
\vfill

\newsec{Introduction}

The two dual descriptions of D-branes as
gravitational solitons of supergravity theories and as objects on which
open strings can end make it possible to derive interesting new relations
between gauge theories and gravity/string-theory. A celebrated example of
this is the Maldacena conjecture \malda . To generalize the
Maldacena conjecture to cases with more complicated gauge groups and
matter content, as well as to cases with less supersymmetry, it is useful
to study D-branes sitting on various spacetime singularities
\refs{\kasi,\lnv,\ansar,\igored}.

This method was pioneered in \dm, where D-branes on orbifold
singularities of the type $C^2/ Z_{k} $ were studied. It was shown
that the gauge theory realized on the brane is conveniently summarized
by a ``quiver diagram'' from which one can read off the gauge group
structure and the matter content. The orbifolds studied in that paper
gave rise to quiver diagrams corresponding to the (extended) $A_{k}$
Dynkin diagram; the generalization to the $D$ and $E$ series,
corresponding to non-abelian orbifolds, was given in \jm . In this
picture, the Fayet-Iliopoulos terms in the gauge theory are related to
twisted sector moduli of the orbifold, so that turning them on
corresponds to blowing up the orbifold singularity.

More complicated orbifolds preserving less supersymmetry were studied
in, \eg, \refs{\gr,\bdual}.

A complementary picture of the same models can be found using
T-duality: the singularity generically transforms into a
web of NS-branes \ogivaf\ and the D-branes change their dimension
to give configurations of the Hanany-Witten type \hw. In this dual
picture, many features of the model become geometric. In particular, the
Fayet-Iliopoulos terms are interpreted as various distances between
branes. In many ways this dual picture is complementary to the
original picture in that different things become easy to see whereas
others, simple in the original picture, become harder to deal
with in the T-dual version of the model. This T-duality has been
studied in detail for the A-series \adual\ and the D-series
\refs{\ddual,\dddual} but no duals have yet been found for the
E-series.

One aim of this paper is to provide guidance to finding such
a correspondence by constructing the map between the Fayet-Iliopoulos
parameters, corresponding to the positions of various branes, and the
deformations of the curve. In this paper, we find the algebraic curve
corresponding to any manifold that is a \hk quotient \refs{\lr,\hklr} of
a linear space. Such a quotient may be described in terms of a quiver
diagram \dm. The cases when it yields a four-dimensional ALE manifold
have been analyzed and have an A-D-E classifcation \refs{\nh,\pk}. In
particular we include the Fayet-Iliopoulos parameters in the calculation
of the curve for the $A_k$, $D_k$ and $E_6$ cases. Remarkably we find that
the curve in the $E_6$ case is identical to the Seiberg-Witten curve
for certain $N=2$ superconformal Yang-Mills theories with $E_6$ global
symmetry and with the Fayet-Iliopoulos terms playing the roles of
chiral superfield VEV's. It would be interesting if some duality or
mirror symmetry were responsible for this apparent coincidence.

The paper is organized as follows.
In section 2 we review the \hk quotient in $N=1$
superspace \refs{\lr,\hklr}. In section 3, we
describe the algebraic curves of these spaces, and derive
them from the $N=1$ superspace description of the
\hk quotient for the $A_k$ \abfgz, $D_k$ \gns, and $E_k$ cases.
In section 4, we discuss a number of issues and consider some higher
dimensional examples outside the A-D-E classification.

\newsec{Hyperk\"ahler quotients}
The \hk quotient was introduced in \lr\ and was given a full
mathematically rigorous presentation in \hklr.
It arises naturally when one gauges isometries of a nonlinear
$\sigma$-model \chklr\ in such a way as to preserve four
dimensional $N=2$ supersymmetry. In components ($N=0$), it is the
quotient of a constrained submanifold (the zero-set of the moment map
\hklr) by some real compact gauge group. In $N=1$ superspace, the
vector multiplet relaxes a part of the constraints, leaving only a
holomorphic constraint, and enhances the gauge group to its
complexification (subtleties pertaining to quotients by noncompact
groups are discussed in \hklr, p.\ 548).

Explicitly, we want to consider the \hk quotient construction of
4-dimensional ALE spaces \pk. We start with a quaternionic vector
space that we describe as an even dimensional complex space with $n$
pairs of coordinates $(z_+,z_-)$. In the language of
supersymmetry, each pair of complex coordinates is called a
hypermultiplet, and in $N=1$ superspace, these are pairs of chiral
superfields. The K\"ahler potential of the metric is the superspace
Lagrangian \bz.  The holomorphic moment map constraints take the form
\lr
\eqn\mm{z_+ T_A z_- = \chi(T_A),}
where the $T_A$ are generators of the gauge group (taken to be
hermitian), and $\chi$ is an arbitrary character--\ie, a
linear combination of the $U(1)$ factors of the group:
\eqn\ulmm{\eqalign{z_+ T_A z_- &=\, 0~~\qquad A\notin \hbox{any
$U(1)$}\cr &=\, b_A\qquad A\in\hbox{any $U(1)$}.}}
In superspace, these are called Fayet-Iliopoulos terms.
The K\"ahler potential of the quotient space is found by solving
a set of real equations for the $N=1$ vector superfields $V^A$ \lr:
\eqn\vsol{z_+e^{V^AT_A}T_A\bar z_+-\bar
z_-T_Ae^{-V^AT_A}z_-=\hat\chi(T_A)~,}
and substituting the solution
into the gauged flat space K\"ahler potential of the $z_\pm$'s \lr:
\eqn\kflat{z_+e^{V^AT_A}\bar z_++\bar
z_-e^{-V^AT_A}z_--V^A\hat\chi(T_A)~,}
where $\hat\chi$ is an independent character. The particular choices
of gauge groups and representations are given in \pk; for a review
see \abfgz. A summary is given in table 1:
\midinsert
$$
{\epsfxsize=12cm\epsfbox{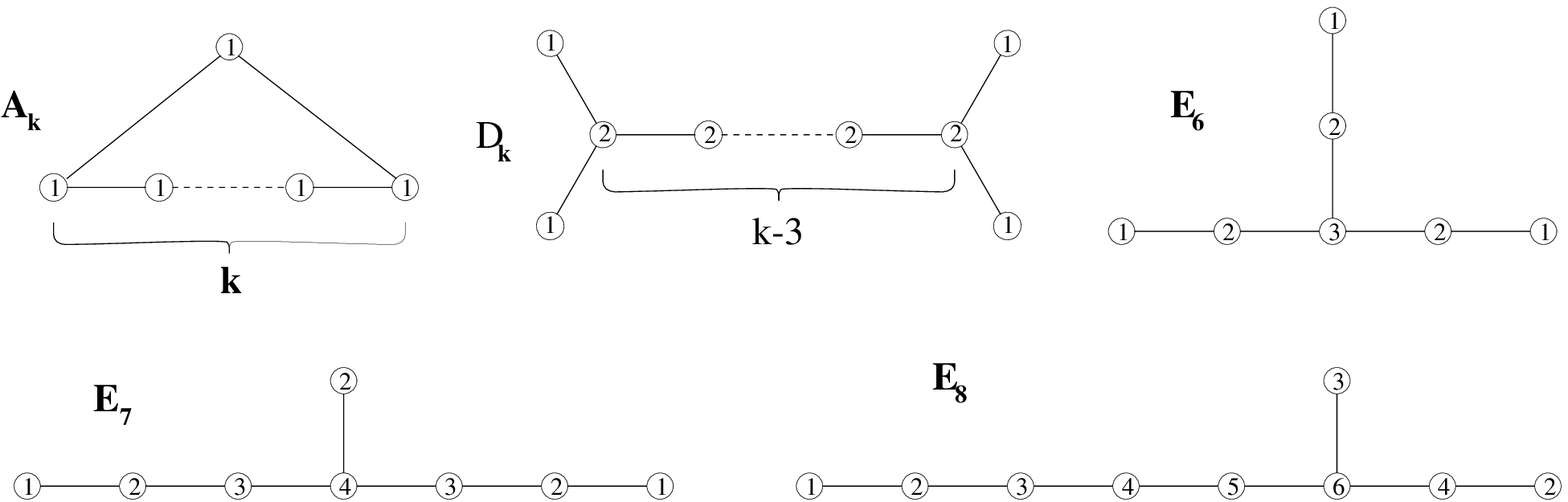}}
$$
\centerline{\bf Table 1}
\endinsert
\noindent In the table, each \hk quotient by a product
gauge group $U(N_1)\times \ldots \times U(N_k)$ is represented as an
extended Dynkin diagram for the A-D-E series of Lie groups; the $i$'th
simple root, which has a label $N_i$ in the Dynkin diagram, corresponds
to a factor $U(N_i)$ in the gauge group and each link between two roots
$i,j$ corresponds to a hypermultiplet in the $(N_i,\bar{N_j})$
representation\foot{As the chiral superfields $z_\pm$
that make up the hypermultiplet are always in conjugate
representations, the orientation of the links does not matter.}.

\newsec{Algebraic curves}
As stated above, four dimensional ALE \hk manifolds are classified by
the extended Dynkin diagrams corresponding to the A-D-E Lie groups
\nh. As complex manifolds, they can be described by holomorphic curves
in ${\bf C}^3$ with coordinates $X,Y,Z$.  The curves have a leading
piece corresponding to the orbifold limit of the spaces, and
deformation parameters corresponding to the character
(Fayet-Iliopoulos) terms of the quotient construction. The curves are
summarized in table 2:

$$
\vbox{\offinterlineskip
\hrule
\halign{&\vrule#&\strut~\hfil#\hfil~\cr
height3pt&\omit&&\omit&&\omit&\cr
&Classification&&Polynomial&&Deformations&\cr
height3pt&\omit&&\omit&&\omit&\cr
\noalign{\hrule}
height3pt&\omit&&\omit&&\omit&\cr
&$A_k$&&$XY-Z^{k+1}$&&$1,\ldots,Z^{k-1}$&\cr
height2pt&\omit&&\omit&&\omit&\cr
&$D_k$&&$X^2+Y^2Z-Z^{k-1}$&&$1,Y,Z,\ldots,Z^{k-2}$&\cr
height2pt&\omit&&\omit&&\omit&\cr
&$E_6$&&$X^2+Y^3-Z^4$&&$1,Y,Z,YZ,Z^2,YZ^2$&\cr
height2pt&\omit&&\omit&&\omit&\cr
&$E_7$&&$X^2+Y^3+YZ^3$&&$1,Y,Y^2,Z,YZ,Z^2,Y^2Z$&\cr
height2pt&\omit&&\omit&&\omit&\cr
&$E_8$&&$X^2+Y^3+Z^5$&&$1,Y,Z,YZ,Z^2,Z^3,YZ^2,YZ^3$&\cr
height2pt&\omit&&\omit&&\omit&\cr}
\hrule}
$$
\vskip -2mm
\centerline{\bf Table 2}
\vskip 3mm

The curve corresponding to a given \hk quotient can be constructed by
finding all the (gauge group) invariant holomorphic polynomials modulo
the holomorphic constraints \ulmm; for the A-D-E spaces, we find exactly
three polynomials that satisfy an algebraic constraint, which, after
suitable redefinitions is precisely the equation of the corresponding
algebraic curve. The construction is in the spirit of Seiberg's
construction of effective superpotentials \ns.

We now describe the actual calculations. To find the correct variables
and derive the curve it is useful to employ the following graphic ``bug
calculus''.

As described above, we can represent any \hk quotient by a product
gauge group $U(N_1)\times \ldots \times U(N_k)$ as a quiver
diagram,\foot{A quiver diagram is essentially a Dynkin
diagram with arrows on the links indicating an orientation; when we
construct invariant polynomials, we need to keep track of the
orientation.} where the $i$'th node, labeled $N_i$, corresponds to
a $U(N_i)$ gauge group and each link (including orientation) between
two nodes $i,j$ corresponds to a hypermultiplet in the
$(N_i,\bar{N_j})$ representation. It is therefore natural to
represent any invariant that can be obtained by multiplying the chiral
fields of the model as a closed oriented loop in the quiver diagram.

The holomorphic constraints \ulmm\ can be
represented in bug calculus. Each gauge group (\ie, node) has its own
constraint. For an ``endpoint'' the constraint is shown in Figure 1a
and for a point in a chain the constraint is shown in Figure 1b.
For more complicated junctions one generalizes this keeping in mind
that the sign of each term is determined by the orientation of the
link it sits on. This is shown for two of the junctions that
appear in this paper: the three and four-point vertices in Figures 1c
and 1d.
\midinsert
$$
{\epsfxsize=12cm\epsfbox{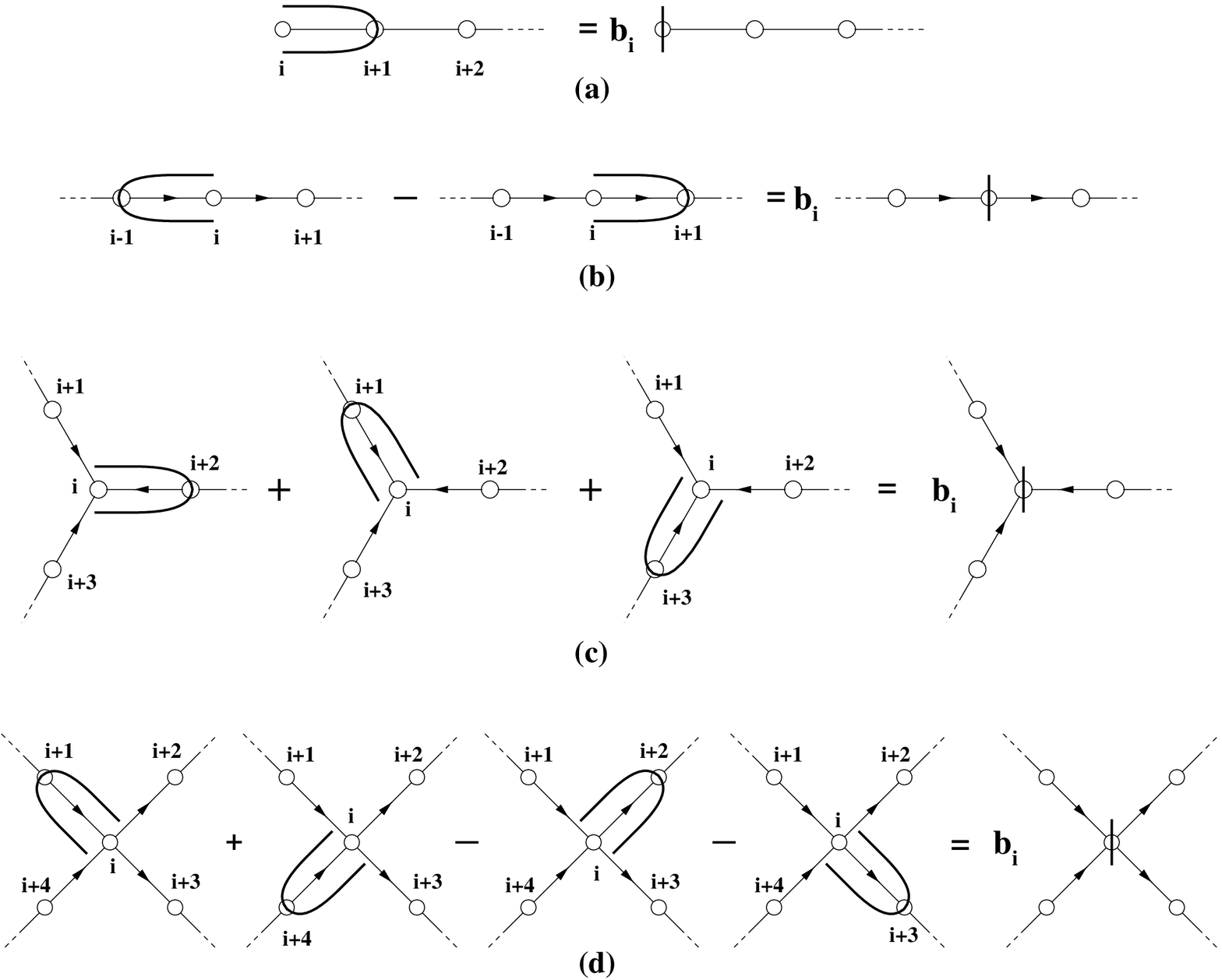}}
$$
\noindent{\bf Figure 1}: The bug calculus. $b_i$ is the Fayet-Iliopoulos
parameter associated to the $i$'th node, and a vertical bar through the
$i$'th node represents a $U(N_i)$ Kronecker-$\delta$.
\endinsert

A second important ingredient is the so-called Schouten
identity. This allows us to untwist various loops showing that
complicated variables can be written as products of simpler ones.
They are derived by observing that the totally antisymmetric product
of $n+1$ $n$-dimensional indices is identically zero. The simplest
identity holds for one dimensional matrices and simply says that one
dimensional matrices commute. Graphically this means that on any node
representing a $U(1)$ gauge group we are allowed to split the lines
and reconnect them in any way as long as we respect the orientation of
the loops.

The Schouten identity for two dimensional matrices looks
slightly more complicated. It can be derived from
\eqn\sid{M^{[i_{1}}_{k_{1}} N^{i_{2}}_{k_{2}} K^{i_{3}]}_{k_{3}} = 0.}
If we contract the indices we can derive the following identity
appropriate for our purposes
\eqn\sch{
 \Tr (\{M,N\} K) = \Tr (MN) \Tr (K)  + \Tr(MK) \Tr (N)
                  +\Tr (NK) \Tr (M) - \Tr (M) \Tr (N) \Tr (K).}
In principle we could also implement this identity
graphically. However, in practice it is easier to use it in algebraic
form and then to go on and use the graphic methods on each term
separately.

These are all the tools we need to derive the algebraic curve for
any \hk quotient corresponding to an arbitrary quiver diagram: we
draw closed loops of increasing order in the number
of links, and use the bug calculus to find the independent
nonvanishing invariants. In practice, we first consider the
orbifold limit, as then the relations between the invariants are
simpler; the final calculations away from this limit then follow
precisely the same route, but yield many more terms. The independent
invariants are good coordinates on the moduli space. When we find no
new independent invariants, we have all the coordinates of the moduli
space. To find the algebraic curve, we consider the product of the
highest order invariant with its orientation reversed image and use
Schouten identities to express it as a product
of lower invariants; for the $D_k$ and $E_k$ but not the $A_k$ cases,
the orientation reversed loop is proportional to the original
invariant plus algebraic functions of the lower invariants. For
the $A_k$ and $D_k$ cases, the $U(1)$
Schouten identity is all we need.
\midinsert
$$
{\epsfxsize=12cm\epsfbox{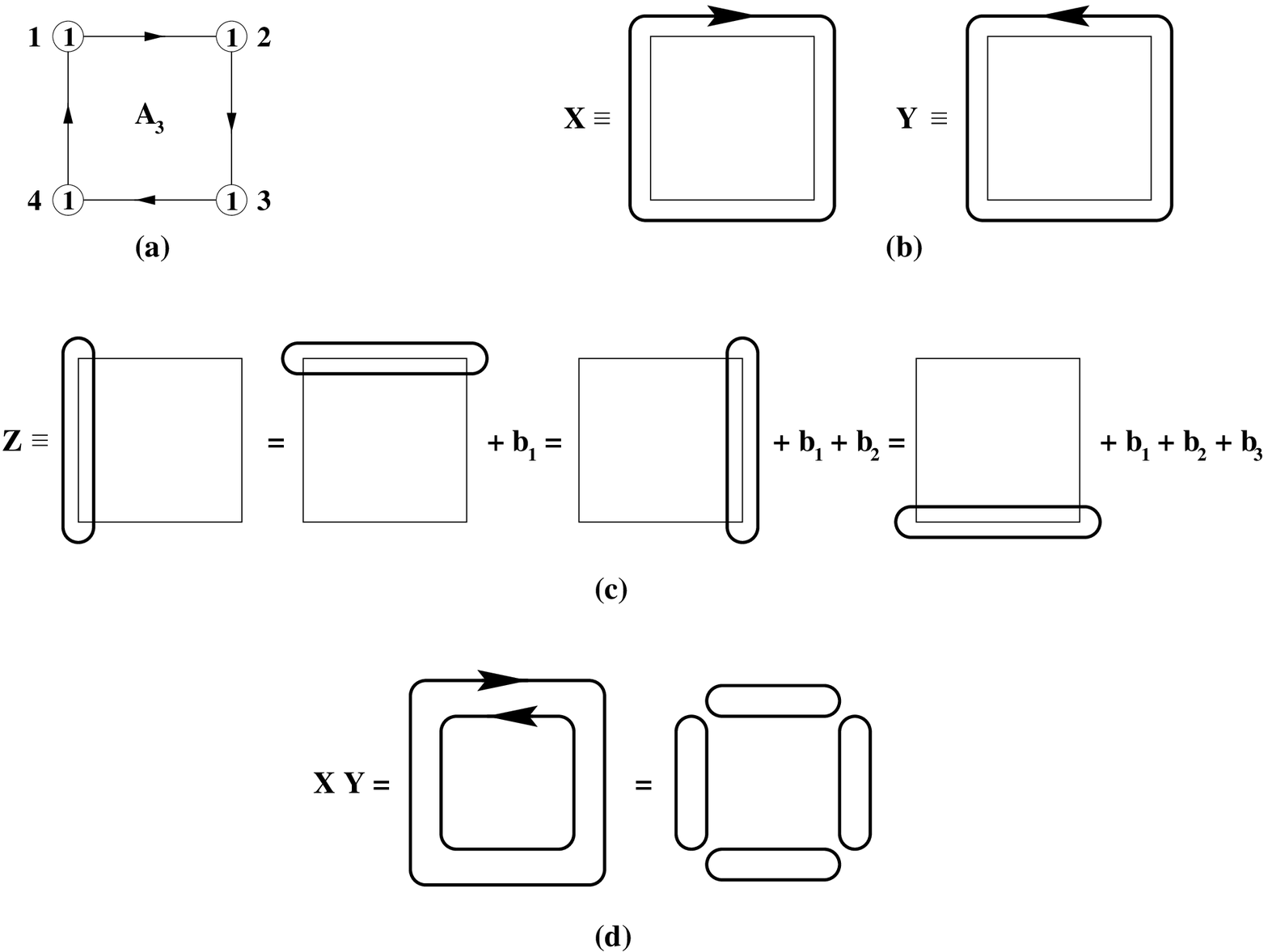}}
$$
\noindent{\bf Figure 2}: Diagramatic representation of the $A_3$
invariants, moment map constraints, and algebraic curve.
\endinsert
We illustrate the method with two examples: $A_3$ and $D_4$, and then
describe our results for the general case. Figure 2 describes the full
calculation for $A_3$: 2a) shows the quiver diagram, 2b) defines two
of the independent variables, $X,Y$, 2c) defines the variable $Z$ and
uses the relation 1b) to express other similar diagrams in terms of
it (note that the Fayet-Iliopoulos terms satisfy $\sum_1^4b_i=0$), and
finally, using the relation in 2c), 2d) gives the algebraic curve in
diagramatic form.\foot{This can be made to match the curve
given in table 2 by shifting $Z\to Z+\nfr14(3b_1+2b_2+b_3)$.}
\eqn\acurve{XY=Z(Z-b_1)(Z-b_1-b_2)(Z-b_1-b_2-b_3)}
The calculation for the general $A_k$ is completely
analogous, and gives the curve $XY=\prod_{i=0}^k
(Z-\sum_{j=1}^ib_j)$, where again $\sum_1^{k+1}b_i=0$.
\midinsert
$$
{\epsfxsize=12cm\epsfbox{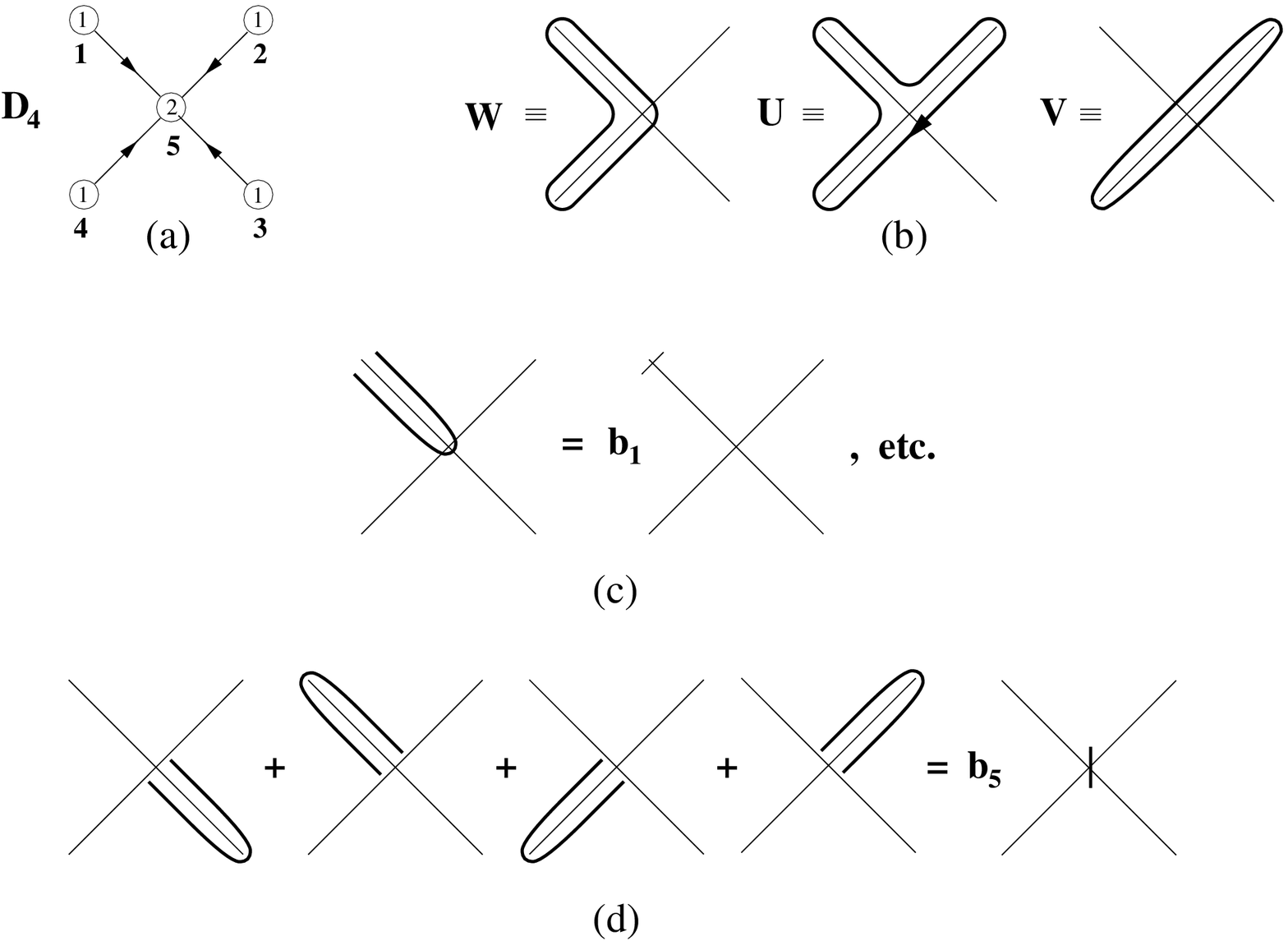}}
$$
\centerline{{\bf Figure 3}: Diagramatic representation of the $D_4$
invariants and moment map constraints.}
\endinsert
Figures 3 and 4 describe the calculation for $D_4$. Figure 3a) shows
the quiver diagram, Figure 3b) defines the three independent variables
and Figures 3c) and 3d) give the constraints 1c) and 1d) for this
particular case; as for the $A_k$ case, the Fayet-Iliopoulos
coefficients are constrained:
$\sum_1^4b_i=2b_5$.
\midinsert
$$
{\epsfxsize=12cm\epsfbox{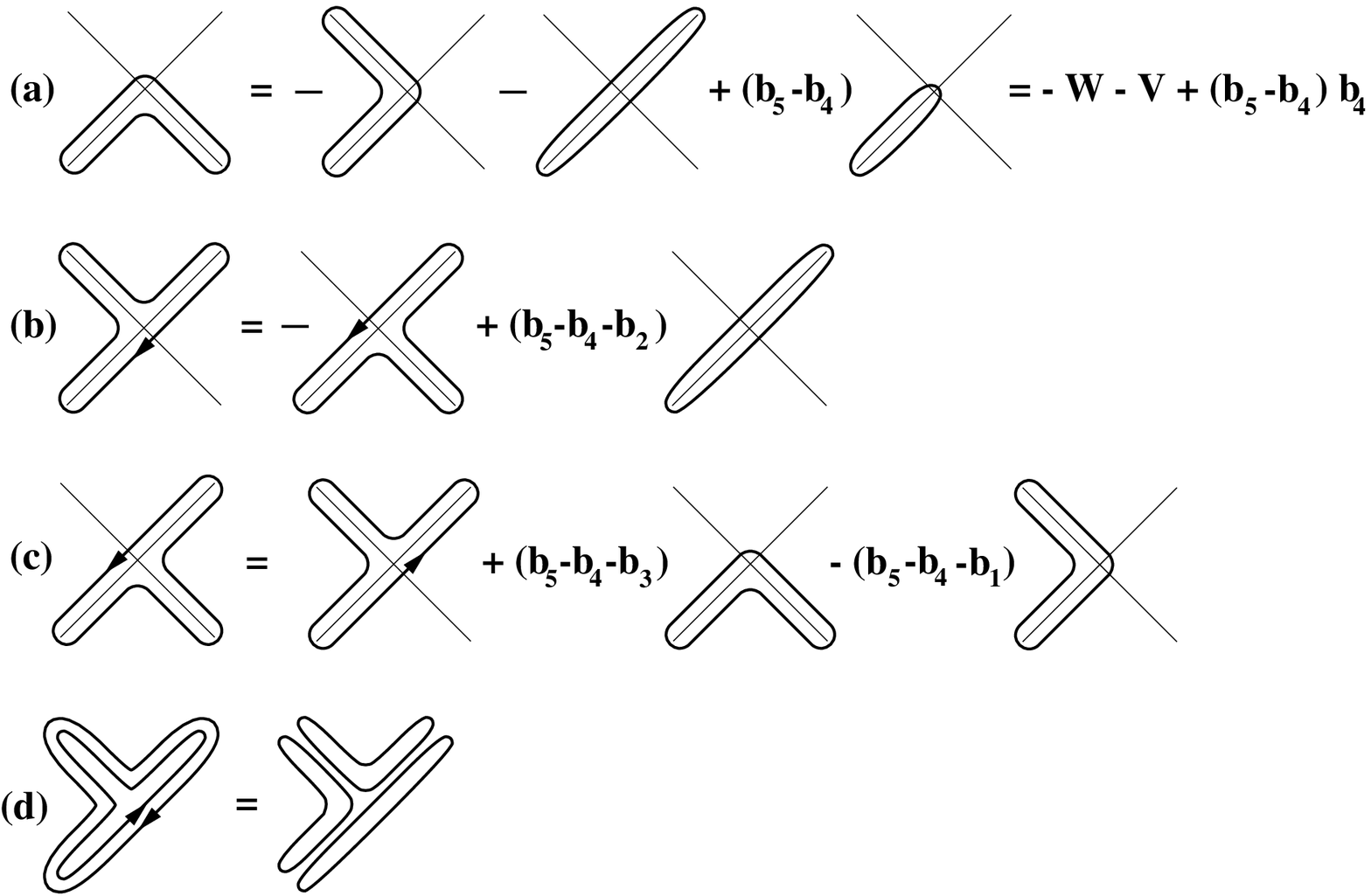}}
$$
\centerline{{\bf Figure 4}: Some typcial calculations for the $D_4$
example.}
\endinsert
Figure 4a)
expresses a four-link diagram in terms of the basic four-link diagrams
$W$ and $V$. Figures 4b) and 4c) relate $U$ to its orientation reversed
image. Figure 4d) yields the algebraic curve in diagramatic form.
Substituting 4a)-c) and similar relations into 4d) we find
\eqn\dcurve{U^2 +
U[(b_4-b_1)V+(b_4-b_2)W+a_1]-W^2V-WV^2+a_2WV=0\ , }
where
\eqn\as{a_1\equiv b_4(b_5-b_4)(b_5-b_4-b_3)~, \quad a_2\equiv
\frac12 \[\sum_{i\ne 3}^4 b_i(b_5-b_i)-b_3(b_5-b_3)\]\ .
\quad}
Making the following redefinitions,
\eqn\red{\eqalign{U&=~ \hf[X+(b_1-b_4)V+(b_2-b_4)W-a_1]~, \cr
V&=~\hf[Y-W+a_2-\hf (b_1-b_4)(b_2-\hf (b_1+b_4))]~,\cr
W&=~ -Z-\nfr14(b_1-b_4)^2~,
}}
we find a standard form of the algebraic curve for $D_4$:
\eqn\algd{
X^2+Y^2Z-Z^3 +\a _0Y-\sum_1^3\a_iZ^{i-1}=0~.}
The coefficients $\a_i$ are expressed in terms of the
Fayet-Iliopoulos  parameters $b_i$ as follows:
\eqn\adefs{\eqalign{\a_0 & =~ \nfr18 (b^2_1-b^2_4) (b^2_2-b^2_3)
~,
\cr
\a_1 &
=\nfr1{32}\[(b_1^2+b_4^2)(b^2_2-b_3^2)^2+(b_2^2+b_3^2)(b_1^2-b_4^2)^2\]
~,
\cr
\a_2 & =~
\nfr1{16}\[(b_1^2-b_4^2)^2+(b_2^2-b_3^2)^2+4(b_1^2+b_4^2)
(b_2^2+b_3^2)\]
~,
\cr
\a_3 & =~\hf\sum_1^4b^2_i~.}}

\midinsert
$$
{\epsfxsize=12cm\epsfbox{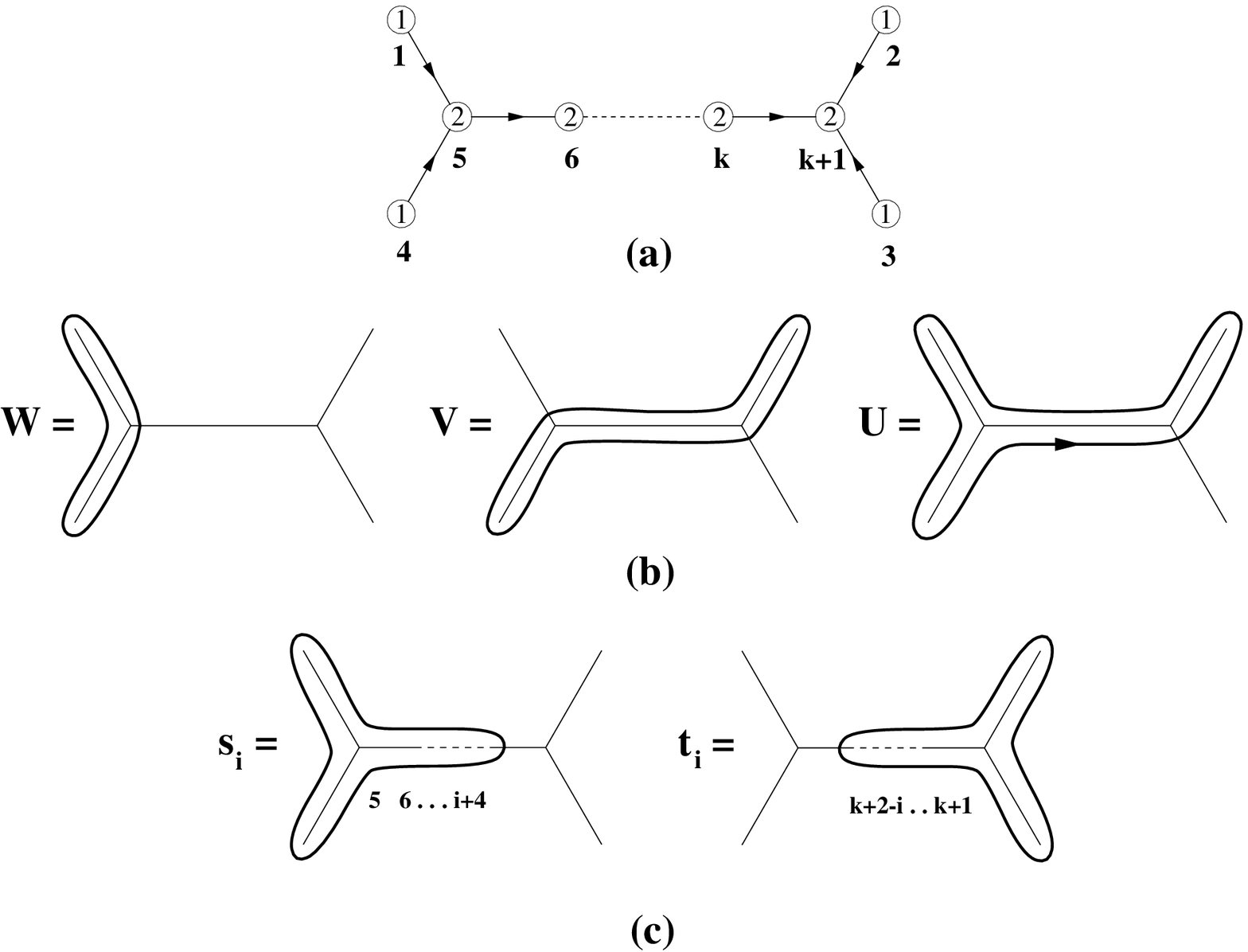}}
$$
\centerline{{\bf Figure 5}: The basic invariants for $D_k$ and some
other useful invariant quantities.}
\endinsert
For the general $D_k$, we label the nodes as indicated in Figure 5a);
the basic variables $U,V,W$ are defined by analogy to the $D_4$ case
(Figure 3b), and are shown in Figure 5b). The Fayet-Iliopoulos
coefficients $b_i$ (associated to the $i$'th node) satisfy
the constraint
\eqn\ficon{\sum_1^4b_i=2\sum_5^{k+1}b_i~.}
We express the curve in terms of certain polynomials in $W$ that we
define recursively as follows:
\eqn\recurs{\eqalign{
s_i&=\[b_1+b_4-b_{i+3}-2\beta_{i-2}\]s_{i-1}+
\[s_1-(b_1-\beta_{i-2})(b_4-\beta_{i-2})\]s_{i-2}~,\cr
t_i&=\[b_1+b_4+b_{k+3 -i}-2\beta_{k-1-i}\]t_{i-1}
+\[s_1-(b_1-\beta_{k-1-i})(b_4-\beta_{k-1-i})\]t_{i-2}
~,\cr
\beta_i&\equiv\sum_{j=5}^{i+4} b_j
~;}}
their graphical expression is given in Figure 5c). The initial
conditions are
\eqn\init{s_0=t_0=0~,~~s_1=W~,~~
t_1=W+\beta_{k-3}(\beta_{k-3}-b_1-b_4)+
\hf(b_1^2+b_4^2-b_2^2-b_3^2)~.}
For arbitrary $k$, the curve in the form analogous to \dcurve\ is
\eqn\dkcurve{U^2-WV^2+U[P(W)+(b_4-b_1)V]+VQ(W)=0~,}
where $P$ and $Q$ are polynomials of order $\OO(W^{[\nfr{k-3}2]+1})$
and $\OO(W^{[\nfr{k-2}2]+1})$, respectively:
\eqn\pqdef{\eqalign{P&=~\frac{b_3-b_2}{t_1}(t_{k-1}-b_1t_{k-2})
+s_{k-2}-\frac{b_1+b_3-\beta_{k-3}}{s_1}[s_{k-1}-
(b_1-\beta_{k-3})s_{k-2}]
~,\cr Q&=~-\frac{s_1}{t_1}[t_{k-1}+(b_3-\beta_{k-3})t_{k-2}]  ~.}}
Though they do not have a graphical representation, $s_i,t_i$ for
$i>k-3$ are defined by the recursion relations \recurs; we also need
$b_i=0$ for $i>k+1$, $\beta_i=0$ for $i<1$, and we take $b_{k+3-i}=0$,
and not $b_4$, for $i=k-1$.  Because of the initial conditions \init,
$s_i/s_1$ and $t_i/t_1$ are polynomials in $W$.

The curve \dkcurve\ can be put into the standard form by redefinitions
analogous to \red:
\eqn\dkred{\eqalign{U&=~ \hf[X+(b_1-b_4)V-P(W)]~, \cr
V&=~ \hf\[Y-\frac{R(Z)-R(0)}Z\]~,\cr
W&=~ -Z-\nfr14(b_1-b_4)^2~,\cr
R(Z)&\equiv \[Q(W(Z))+\hf(b_1-b_4)P(W(Z))\]_{W(Z)=~
-Z-\nfr14(b_1-b_4)^2}~,}}
which gives
\eqn\dkstan{X^2+Y^2Z+2R(0)Y-
\left[\frac{R^2(Z)-R^2(0)}Z+P^2(W(Z))\right]=0~.}
Calculating the first six examples,
we are able to rewrite the curve \dkstan\ explicitly in terms of
the Fayet-Iliopoulos coefficients:
\eqn\dkexp{X^2+Y^2Z-2Y\prod_1^kB_i-
\frac{\prod_1^k(Z+B^2_i)-\prod_1^kB^2_i}{Z}=0~,}
where
\eqn\Bdef{
\{B_i\}\equiv \left\{\hf (b_1-b_4),\hf (b_2-b_3),\hf (b_1+b_4),
\hf (b_1+b_4)-b_5,\dots ,\hf
(b_1+b_4)-\sum_5^{k+1}b_i\right\}~.}
In the orbifold limit, $B_i=0$ which agrees with the entry for
$D_k$ in table 2. After completing our calculation, we realized that
the same expression for the deformation in terms of the
Fayet-Iliopoulos parameters had been deduced by completely different
methods in \gns.

We note that the quantities that enter in both
the $A_k$ and the $D_k$ cases are related to the weights of the
fundamental representation of the Lie algebra in question. If we
think of each Fayet-Iliopoulos parameter as the simple root associated
to its node in the Dynkin diagram, then the expressions that occur
($\sum_{j=1}^{i} b_{j}$ in the $A_k$ case and the $B_{i}$ \Bdef\ in the
$D_{k}$ case) are the weights of the fundamental representation.
More precisely, since the Fayet-Iliopoulos terms are scalars
whose value may be freely chosen, the quantities entering the curve
should be associated with $v\cdot\lambda$ where $\lambda$ is the
particular weight and $v$ is a vector of the same dimension as the
rank of the group. This ensures that we can choose the
Fayet-Iliopoulos parameters to be zero, corresponding to a zero value
for $v$ and when we turn on the Fayet-Iliopoulos parameters it
corresponds to giving $v$ a non-zero value such that the quantities above
agree. This observation will be used later to write the result for the
$E_6$ curve in a nice form.

\midinsert
$$
{\epsfxsize=12cm\epsfbox{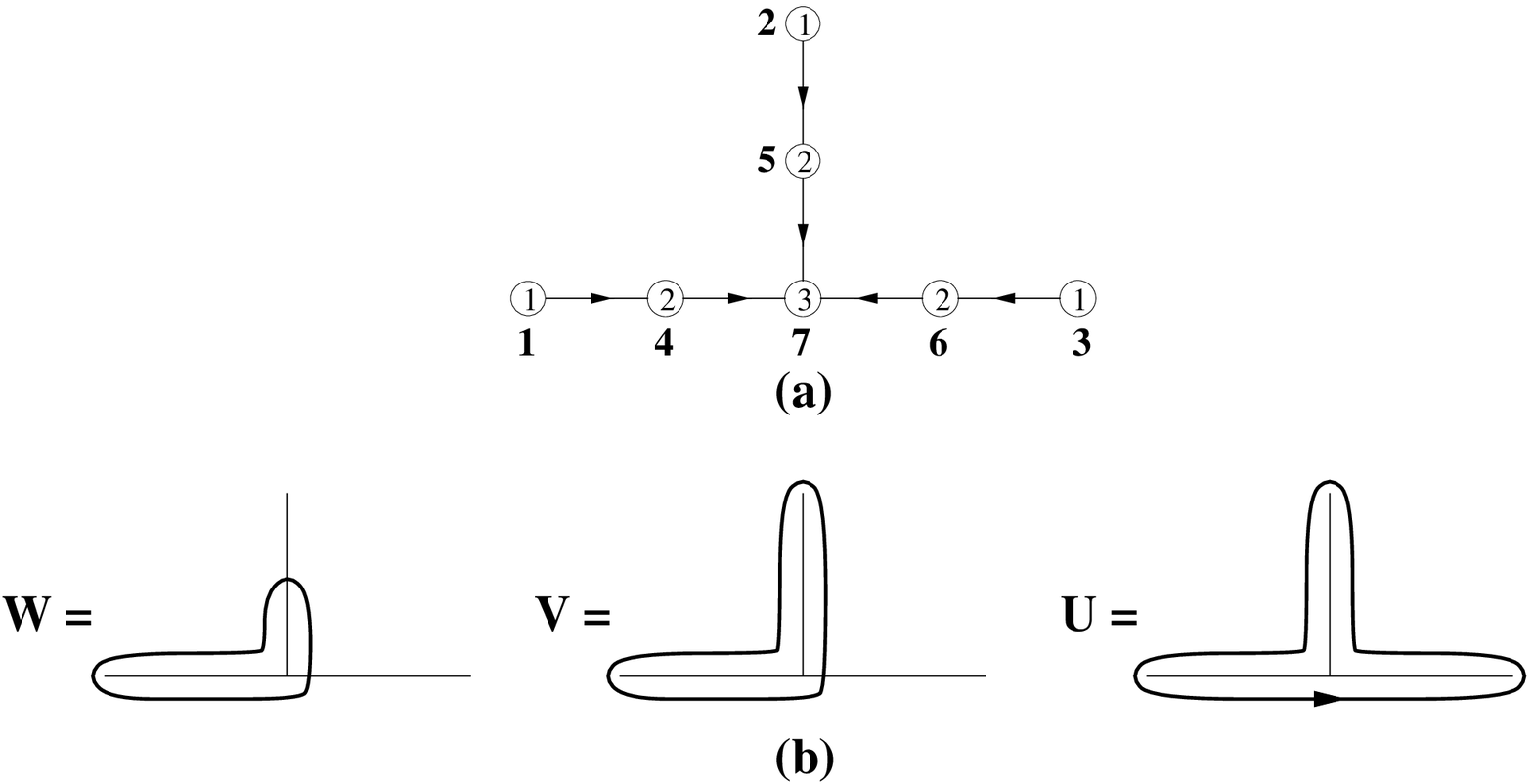}}
$$
\centerline{{\bf Figure 6}: The $E_6$ invariants.}
\endinsert
We now turn to the $E$-series.
The labeling of the nodes for $E_6$ is given in Figure 6a), and the
invariant polynomials $U,V$ and $W$ are defined in Figure 6b). The
relation between $U$ and its orientation reversed image $\bar U$ is:
\eqn\esexu{
U+\bar U= -W^2+A_WW+A_VV+A_0 ~,}
where
\eqn\Awvo{\eqalign{
A_V&\equiv\sum_1^3 b_i(b_i-b_{i+3})-\nfr14(\sum_1^3b_i+b_7)^2
~,\cr
A_W&\equiv
\nfr14(\sum_1^3b_i+b_7)(b_4+b_5-b_1-b_2+b_7)(b_1+b_2-b_3+b_7)\cr
&-b_1b_5(b_1-b_4)+b_2(2b_1-b_4)(b_2-b_5)\cr
A_0&\equiv -b_2(b_1+b_3-b_4-b_6-b_7)C_0~,\cr
C_0&\equiv b_1(b_1-b_4)(b_1-b_4-b_7)(b_1-b_4-b_6-b_7)~.
}}
Just as in the $D_4$ case (\cf\ Figure 4d) the curve follows from
expressing $U^2$ as $U(-\bar U+...)$. The result is
\eqn\esix{\eqalign{&U^2-U(-W^2+A_WW+A_VV+A_0)\cr
&+V\left[V+C_WW+C_0\right][V+D_WW+D_0] =0
}}
where
\eqn\cdef{\eqalign{
C_W&\equiv b_1-b_2-b_3+b_5+b_6+b_7~,\cr
D_W&\equiv b_1-b_2+b_3-b_4-b_6-b_7~,\cr
D_0&\equiv \nfr{1}{27}\[2b_1^3
-\nfr{27}{8}(\nfr{1}{3}b_1-b_2+b_3-b_7)(\nfr{1}{3}b_1+b_2-b_3-b_7)
(\nfr{1}{3}b_1-b_2+b_3-2b_6-b_7)\cr
&
+3b_1\left\{(2b_1-b_2-2b_4+b_5)(b_1-2b_2-b_4+2b_5)
-2\left(b_6^2+(b_2-b_3)(b_2-b_3+b_6)\right)\right\}\]\cr
&\times(b_1+b_3-b_4-b_6-b_7)~.
}}
Performing the following shifts
\eqn\esshift{\eqalign{U&=X+\nfr{1}{2}(-W^2+A_WW+A_VV+A_0)~,\cr
V&=Y-\nfr{1}{3}\left[ C_0+D_0+(C_W+D_W)W-A^2_V/4\right]~,\cr
W&=\sqrt{2}Z+\nfr{1}{6}\left[3A_W-
\left(A_V+\nfr{2}{3}C_WD_W\right)(C_W+D_W)+\nfr{4}{9}
\left(C_W^3+D_W^3\right)\right]~,}}
the curve \esix\ is brought to the standard form
\eqn\esixs{X^2+Y^3-Z^4+P(Z)+Q(Z)Y=0~,}
where the polynomials $P(Z)$ and $Q(Z)$ are second order in $Z$.
The coefficients in terms of the Fayet-Iliopoulos
parameters may be found by substituting \esexu, \Awvo, \cdef, and
\esshift\ into \esix. Direct evaluation leads to a horrible mess,
but the polynomials may be expressed in terms of Casimir operators of
$E_6$; remarkably, when we do this, we find the algebraic curve
given in \mn. We now present the details of this description.

The Casimirs can be defined as the coefficients of the polynomial
\eqn\cas{\det \left( x - \Phi \right)~,}
where $\Phi$ is a matrix in the fundamental representation of
$E_6$. We can always rotate $\Phi$ into some element in the Cartan
subalgebra $v\cdot H$ where $v$ is an arbitrary six dimensional vector.
An explicit representation for the matrices $H$ can be found in terms of
the weights $\lambda$ of the fundamental representation, since the weight
vectors can be thought of as normalized eigenvectors of the Cartan
operators with the weights as eigenvalues. The Cartan operators are thus
represented as diagonal matrices with the particular weights on the
diagonal, and we have
\eqn\cart{
\Phi = v\cdot H =
\pmatrix{v\cdot\lambda_1 & \dots & 0 \cr
\vdots & \ddots & \vdots \cr
0 & \dots & v\cdot\lambda_{27}}.}
The terms on the diagonal are just the expressions for the weights
in terms of the Fayet-Iliopoulos parameters as discussed at the end of
the derivation of the curve for the $D_k$ case. Thus we have found a
way to express the Casimirs in terms of the Fayet-Iliopoulos
parameters. More details as well as the final result for the curve can
be found in the appendix.

It is natural to conjecture that the relation between the
deformation parameters of the curve and the Fayet-Iliopoulos
parameters follows the same pattern for the higher exceptional
algebras \refs{\mnII,\nty}. The Fayet-Iliopoulos are to be thought of as
the simple roots of the algebra and the Casimirs of the fundamental
representation of the algebra (expressed in terms of the simple roots
and thus in terms of the Fayet-Iliopoulos paramters) give the
deformation parameters of the curve.

We note that this explicit expression stands in contrast to the
implicit one of \gns, which involves inverting elliptic integrals.

We now turn to the $E_7$ and $E_8$ cases.

\midinsert
$$
{\epsfxsize=12cm\epsfbox{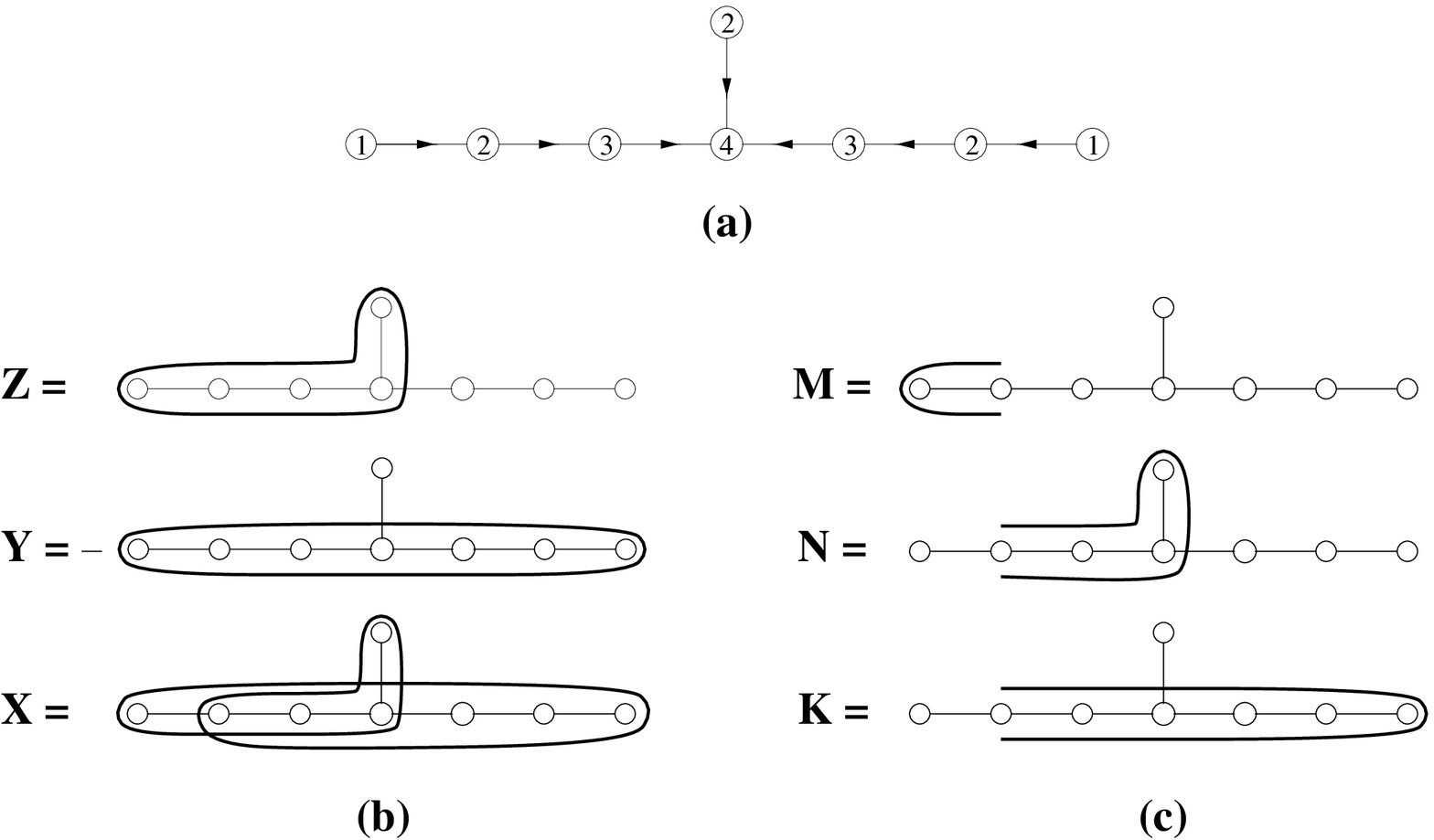}}
$$
\centerline{{\bf Figure 7}: The $E_7$ invariants and some useful
matrices.}
\endinsert
For $E_7$ and $E_8$, we consider only the orbifold limit (no
Fayet-Iliopoulos terms).
In Figure 7, the quiver diagram (Figure 7a) and
the basic invariant polynomials (Figure 7b) for $E_7$ are shown.
We have verified that all other possible invariants either vanish or are
polynomials in these basic variables. In the $E_7$ case the orientation
reversal of the highest dimension graph $X$ is just $-X$, but when we
multiply them together the result does not immediately factorize into a
sum over products of the basic lower dimensional variables. It therefore
turns out to be convenient do define the traceless $2\times 2$ matrices
$M,N$ and $K$ as in Figure 6c. Using the bug calculus it is possible to
derive the following useful relations
\eqn\trcs{\eqalign{
\Tr\left(NK\right) &=~ -Z^2~, \cr
\Tr\left(MK\right) &=~ -Y~, \cr
\Tr\left(MN\right) &=~ Z~, \cr
\Tr\left(N^2\right) &=~ -2Y~.}}
Using these matrices we can write the square of the highest
dimensional invariant as $X^2 = Y \Tr\left(MNKN\right)$. To be able to
use the Schouten identity \sch\ we rewrite the trace in terms of
anticommutators by anticommuting the leftmost matrix all the way to
the right. We can now rewrite the trace in terms of products of traces
of fewer matrices. The result, dropping terms that vanish, is
\eqn\nice{\Tr\left(MNKN\right) = \Tr\left(MN\right)
\Tr\left(KN\right) - \hf \Tr\left(N^2 \right) \Tr \left(MK\right),}
which, using the relations \trcs\ gives the curve
\eqn\eseven{X^2 + Y^3 + Y Z^3 = 0~.}
\midinsert
$$
{\epsfxsize=12cm\epsfbox{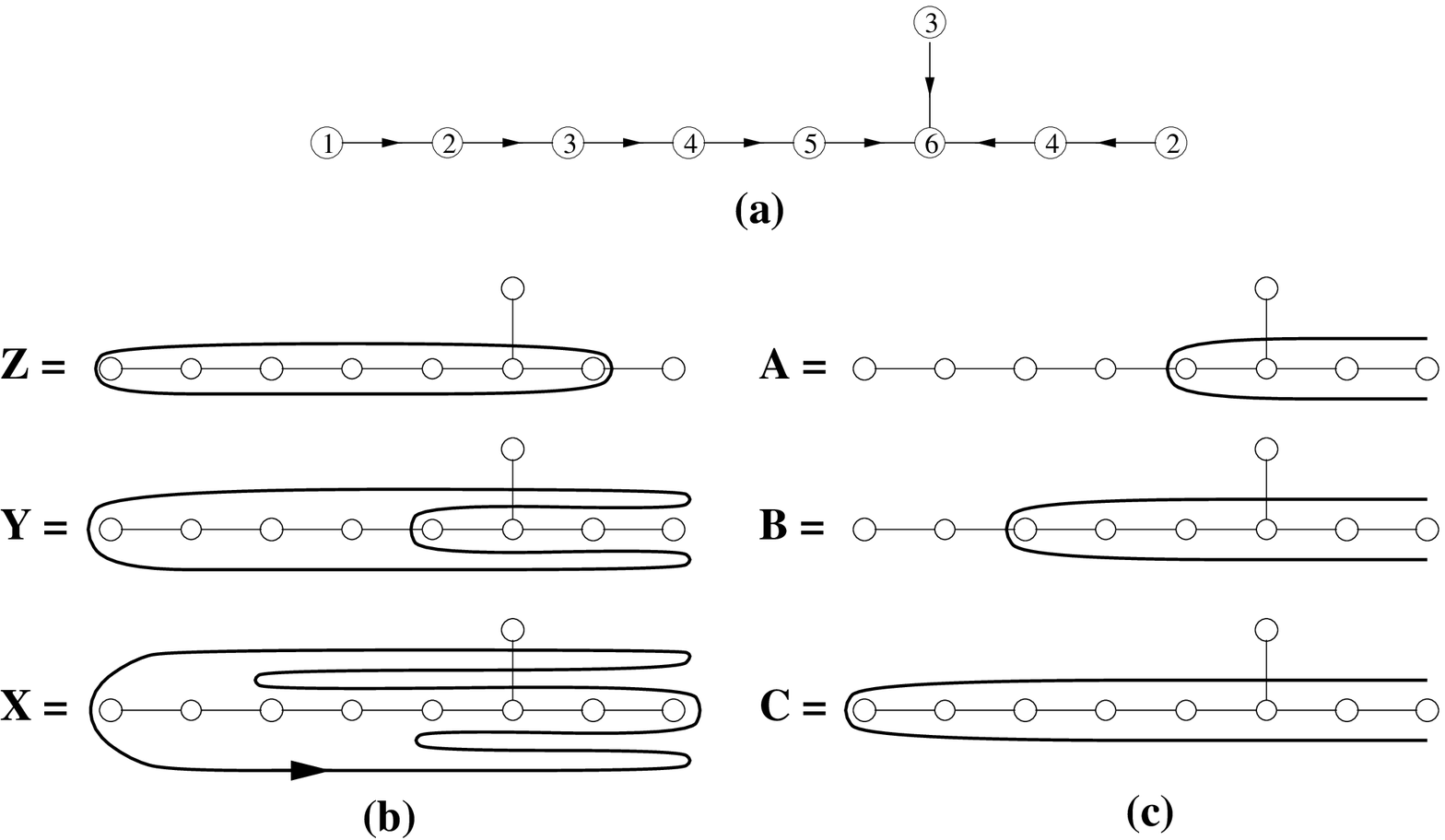}}
$$
\centerline{{\bf Figure 8}: The $E_8$ invariants and some useful
matrices.}
\endinsert
Finally we turn to the $E_8$ ALE space; the quotient gauge group
and matter content are described by the quiver diagram in Figure 8a),
and the basic invariant polynomials are defined in Figure 8b).
Again, we have verified that all other possible invariants either
vanish or are polynomials in these basic variables.
Since there is only one $U(1)$ group in this case it is not
possible to simply factorize the square of the highest dimensional
invariant $X$ into a product of lower dimensional ones and we must again
use the two dimensional Schouten identity \sch. Therefore it is useful
to define the traceless $2\times 2$ matrices $A,B$ and $C$ in Figure
8c). Using the bug calculus we derive the following identities
\eqn\eightraces{\eqalign{
\Tr\left(ABC\right) &=~ X~, \cr
\Tr\left(AB\right) &=~ 0~, \cr
\Tr\left(BC\right) &=~ Z^2~, \cr
\Tr\left(A^2\right) &=~ -2Z~, \cr
\Tr\left(B^2\right) &=~-2Y~.}}
Squaring the highest dimensional invariant $X$ and using the one
dimensional Schouten identity we can write the result as $X^2 =
\Tr\left(ABCABC\right)$. Rewriting the trace in terms of
anticommutators by using the same trick as in the $E_7$ case we get
\eqn\halfthere{
X^2 = \Tr\left(AC\right) \Tr\left(AB^2 C\right) -
    \Tr\left(BC\right) \Tr\left(ACAB\right)~,}
and using the same trick once again on the traces with four matrices
we find
\eqn\fourthere{\eqalign{
\Tr\left(AB^2 C\right) &=~ \hf \Tr\left(AC\right)\Tr\left(B^2
\right), \cr
\Tr\left(ACAB\right) &=~ -\hf \Tr\left(A^2\right)
\Tr\left(BC\right)~.}}
Finally, using \eightraces, we arrive at the following result for
the curve
\eqn\eightcurve{ X^2 + Y^3 + Z^5 = 0~.}

\newsec{Other examples}

There is something a bit surprising about our calculations: aside from
those few nodes where we used the Schouten identities, our calculations
did not in any way refer to the gauge group associated with each node of
the quiver. Thus if we change the Dynkin indices of those nodes where we
did not use a Schouten identity, we get the same invariants and the same
algebraic curve. However, when we consider the \hk quotient, this is
clearly nonsensical: the delicate balance between the dimension of the
gauge group and the number of hypermultiplets is achieved only for the
correct Dynkin indices: \eg, for $D_4$, if the central node is changed
from $U(2)$ to $U(n)$, the resulting \hk quotient has zero or negative
dimension. The resolution of this paradox becomes clear when we express
the fields of the hypermultiplets in ``spherical''-type coordinates,
that is, in terms of goldstone modes that transform under the gauge
group (``angles''), and the invariants (``radii''). When the dimension
of the \hk quotient is zero, the hypermultiplet action depends only on
the goldstone modes, and the invariants that live on the algebraic curve
do not enter the dynamics (one could imagine that under some
circumstances these invariants correspond to dynamically generated
states of the theory, and then the nontrivial \hk quotient manifold
would arise); if the \hk quotient would give rise  to a negative
dimension space, then the hypermultiplet action is not only independent
of the invariants, but even of some of the goldstone modes.

\midinsert
$$
{\epsfxsize=12cm\epsfbox{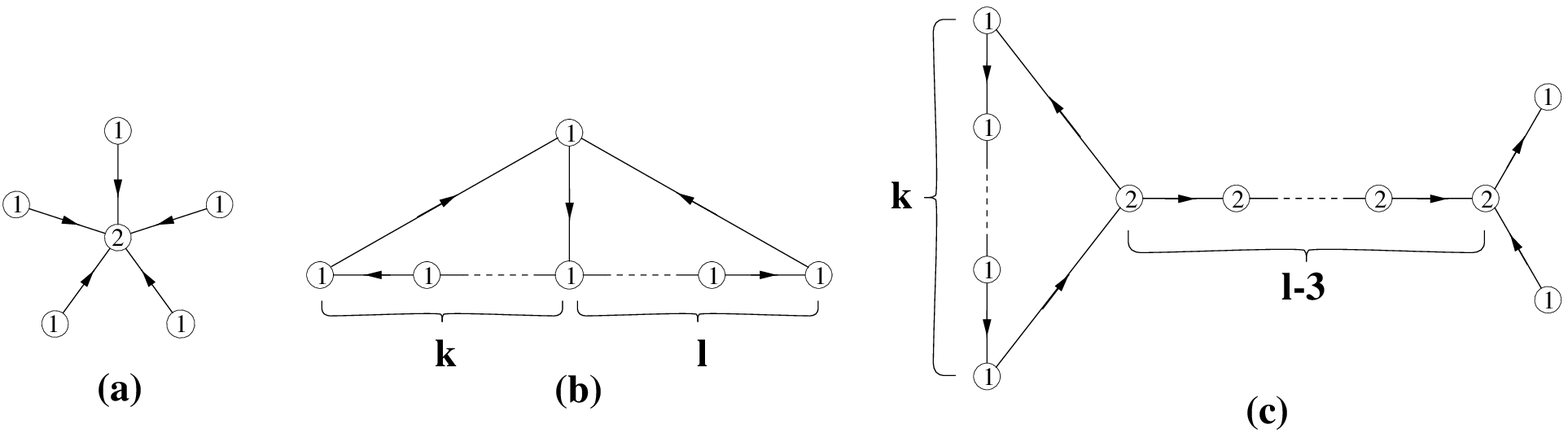}}
$$
\centerline{{\bf Figure 9}: Some higher dimensional examples.}
\endinsert
The graphical methods that we have developed can be used
to find the algebraic curves for higher dimensional ALE spaces. A few
typical examples are shown in Figure 9. We have analyzed only the
orbifold limits of these examples.
\midinsert
$$
{\epsfxsize=13cm\epsfbox{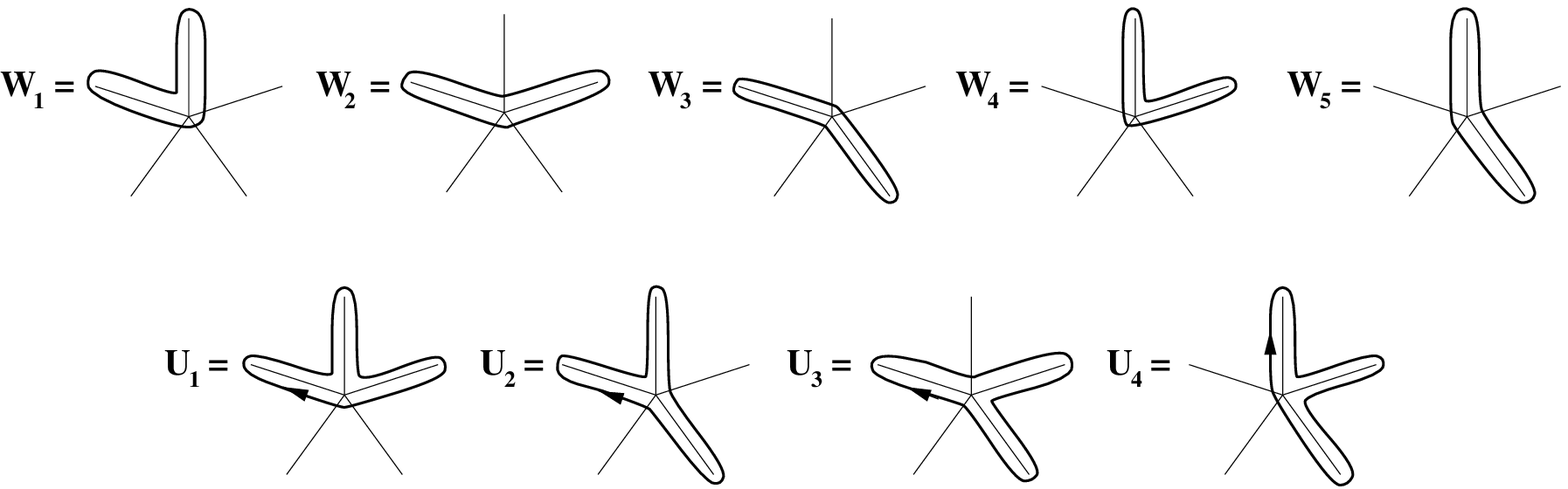}}
$$
\centerline{{\bf Figure 10}: The invariants for an eight-dimensional
analog of $D_4$.}
\endinsert
For Figure 9a), there are nine linearly independent invariants as
defined in Figure 10. Direct application of our method gives ten
polynomial  equations that these invariants satisfy. However, a little
calculation shows that these ten equations are generated by five
equations, leaving a complex four dimensional space as expected from
the \hk quotient:
\eqn\fivestar{\eqalign{U_1^2 =-W_1W_2W_4~,~~ U_2^2 & =-W_2W_3W_5
~,~~ U_3^2=-W_2W_3\sum_1^5W_i~,~~
U_4^2 = -W_4W_5\sum_1^5W_i~,\cr
4\prod_2^5W_i & = (W_2W_5+W_3W_4+W_1\sum_1^5W_i)^2~.}}
This space should be an
interesting nontrivial extension of $D_4$ to higher dimensions. It is
straightforward to find the ten equations with the Fayet-Iliopoulos
terms  turned on; however, in that case, the reduction to five equations
seems to be tedious.
\midinsert
$$
{\epsfxsize=13cm\epsfbox{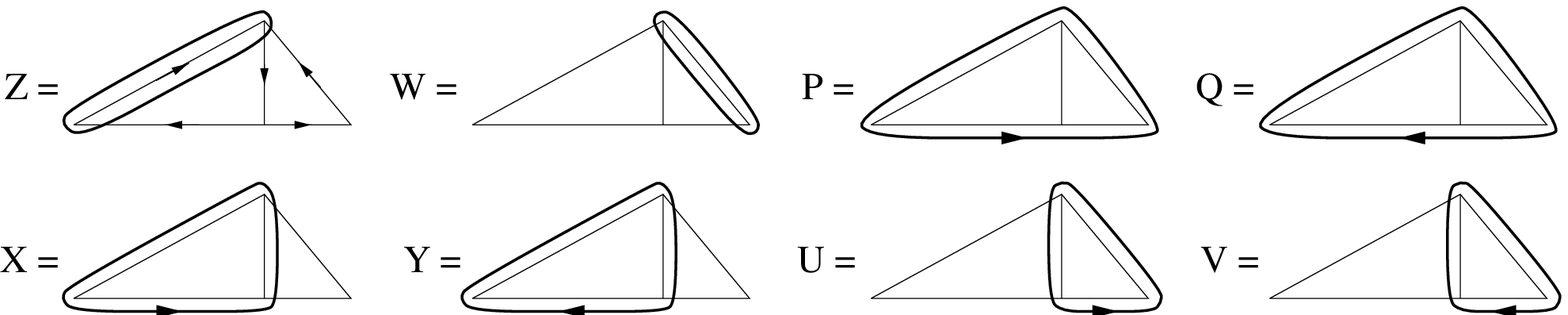}}
$$
\centerline{{\bf Figure 11}: The invariants for eight-dimensional
Calabi ALE spaces.}
\endinsert
For Figure 9b), there are eight linearly independent invariants as
defined in Figure 11. These obey four relations, leaving a complex four
dimensional space:
\eqn\AkAl{XY=Z^k(Z+W)~,~~ UV=W^l(Z+W)~,~~ XU=P(Z+W)~,~~YV=Q(Z+W)~.}
Note that in the orbifold limit, away from the subspace $Z+W=0$, this is
just the product space $A_k\times A_l$.  These spaces are well
understood higher dimensional analogs of the $A_k$ ALE spaces; examples
were constructed as \hk quotients in \lr, though they had been proposed
earlier as \hk spaces in \ec.
\midinsert
$$
{\epsfxsize=13cm\epsfbox{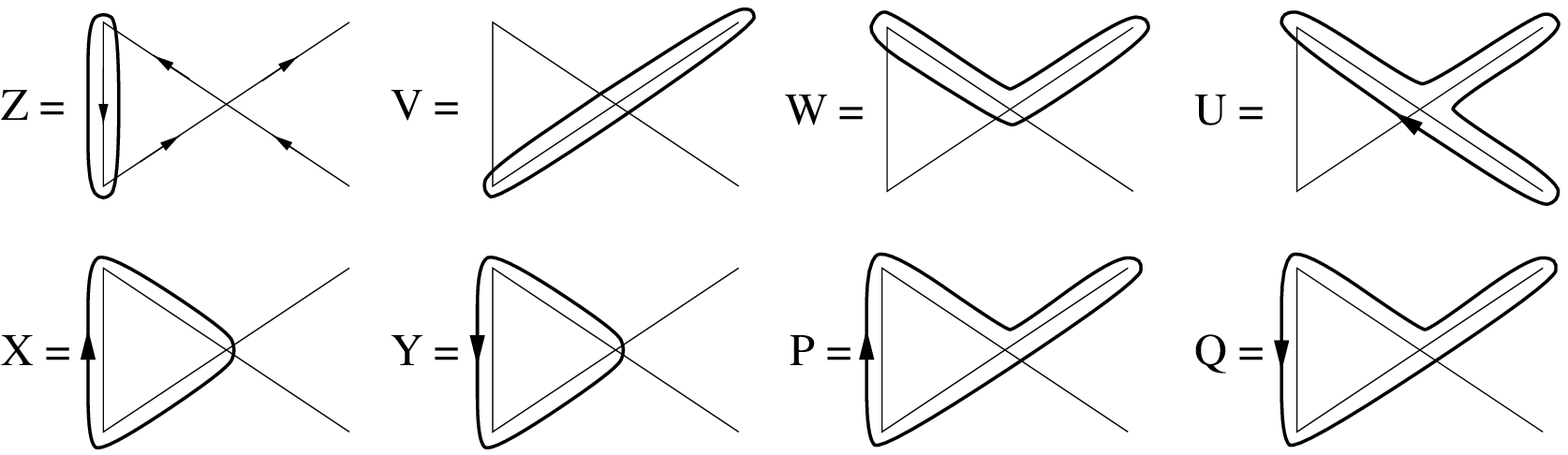}}
$$
\centerline{{\bf Figure 12}: The invariants for another
eight-dimensional example.}
\endinsert
For Figure 9c), there are eight linearly independent invariants as
defined in Figure 12. These obey four relations, leaving a complex four
dimensional space:
\eqn\AtwoDfour{XQ=Z(ZV+U)~,~~ YP=Z(ZW-U)~,~~ XY=Z(Z^2+W-V)~,~~PQ=ZWV~.}

\bigskip
\bigskip
\noindent{\bf Acknowledgements:}
We are happy to thank Gordon Chalmers, Sergei Cherkis, and Blaine Lawson 
for useful discussions. The work of UL was supported in part by NorFA grant
No.\ 9660003-O and NFR grant No.\ F-AA/FU 04038-312. UL would like to
thank the CNY-ITP for hospitality. The work of MR was supported in
part by NSF grant No.\ PHY9722101.  The work of RvU was supported in
part by the Swedish Institute. MR and RvU are happy to thank the
ITP at the University of Stockholm for hospitality.
\vfill
\appendix{A}{Details on $E_6$}
The algebraic curve for the $E_6$ case including the Fayet-Illiopoulos
parameters is
\eqn\finalesix{\eqalign{
X^2 + Y^3 + &\[ \nfr13 P_2 Z^2
 - \nfr23 P_5 Z + \nfr{8}{15}P_8 - \nfr{11}{45}P_2 P_6
 + \nfr{7}{432} P_{2}^{4} \]Y + \cr
 &\[-Z^4-(\nfr{2}{3}P_6 - \nfr{7}{108}P_{2}^{3})Z^{2}+
(\nfr{8}{21}P_9 - \nfr{1}{18}P_{2}^{2}P_5)Z  \cr
 &~-\nfr{32}{135}P_{12} + \nfr{298}{18225}P_{2}^{2}P_8 + \nfr{101}{218700}
  P_{2}^{3}P_{6} - \nfr{13}{405}P_{6}^{2} + \nfr{49}{1049760}P_{2}^{6}
  + \nfr{19}{3645}P_{2}P_{5}^{2}\] = 0}}
where the $P_{i}$ is the Casimir of the $i$'th order. It can be found as
the coefficient of $x^{27-i}$ term of the polynomial
\eqn\casdef{
\det\left(x - v\cdot H \right)
}
where $v\cdot H$ is given in terms of the weights $\lambda $ of the
fundamental representation as $v\cdot H = {\rm diag} \left( v
\cdot \lambda_{1}
\ldots v \cdot \lambda_{27}\right)$. In particular, if we define
$\chi_{n} \equiv \Tr \left[ \left( v\cdot H \right)^n \right]$, we find
that we can write the relevant $P_{i}$ as follows
\eqn\Ps{\eqalign{
P_2 &= -\nfr{1}{2} \chi_2~, \cr
P_5 &= \nfr{1}{5} \chi_5~, \cr
P_6 &= -\nfr{1}{6}\chi_6 - \nfr{1}{96}\chi_{2}^{3}~, \cr
P_8 &= -\nfr{1}{8}\chi_8 + \nfr{1}{12}\chi_2 \chi_6
            +\nfr{1}{4608}\chi_{2}^{4}~, \cr
P_9 &= \nfr{1}{9}\chi_9 - \nfr{1}{14}\chi_2 \chi_7 +
             \nfr{1}{48}\chi_{2}^{2}\chi_{5}~, \cr
P_{12} &= -\nfr{1}{12}\chi_{12}
         +\nfr{11}{1920}\chi_{2}^{2}\chi_{8}
         +\nfr{1}{72}\chi_{6}^{2} - \nfr{1}{1440} \chi_{2}^{3}\chi_6
         +\nfr{17}{2400}\chi_2\chi_{5}^{2}+\nfr{1}{663552}\chi_{2}^{6}~,
}}
where, for $E_6$\foot{After we completed our calculations, we were informed
that such formulas are derived in great generality in \mp.},
\eqn\deptr{\eqalign{
\chi_4 &=~ \nfr{1}{12}\chi_{2}^{2}~, \cr
\chi_7 &=~ \nfr{7}{24}\chi_{2}\chi_5~ , \cr
\chi_{10} &=~ \nfr{7}{40}\chi_{5}^{2} + \nfr{3}{8}\chi_{2}\chi_{8}
  -\nfr{7}{144}\chi_{2}^{2}\chi_{6} + \nfr{7}{41472}\chi_{2}^{5}~.}}
To express the Casimirs in terms of the Fayet-Iliopoulos
parameters, we write the weights of the fundamental representation in
terms of the simple roots and recall that each $b_i$ can be thought
of as the scalar product between the $v$ and its corresponding simple
root.\foot{In these formulas, we have eliminated $b_7$ using
the relation $\sum_1^3b_i+2\sum_4^6b_i+3b_7=0$. It is
straightforward to use this formula to eliminate one of $b_1,b_2,b_3$,
to get the expressions in terms of the more
standard simple roots. We have also switched
the signs of $b_1,b_2,b_3$ as compared to the text
to agree with the usual conventions for the simple roots, which do
not agree with the signs we read off from the quiver diagrams.}
Doing this we find
\eqn\fundweight{\eqalign{
{v\cdot\lambda_{1}} &=~ { \nfr {1}{3}} \,{b_{5}} -
{ \nfr {1}{3}} \,{b_{4}} - { \nfr {2}{3}} \,{b_{1}}
+ { \nfr {2}{3}} \,{b_{2}} \qquad~~\,
{v\cdot\lambda_{2}} ~=~ { \nfr {1}{3}} \,{b_{5}} -
{ \nfr {1}{3}} \,{b_{4}} - { \nfr {2
}{3}} \,{b_{1}} - { \nfr {1}{3}} \,{b_{2}} \cr
{v\cdot\lambda_{3}} &=~  - { \nfr {2}{3}} \,{b_{5}} -
{ \nfr {1}{3}} \,{b_{4}} - { \nfr {2
}{3}} \,{b_{1}} - { \nfr {1}{3}} \,{b_{2}} \qquad
{v\cdot\lambda_{4}} ~=~ { \nfr {1}{3}} \,{b_{4}} -
{ \nfr {1}{3}} \,{b_{1}} + { \nfr {1
}{3}} \,{b_{3}} + { \nfr {2}{3}} \,{b_{6}} \cr
{v\cdot\lambda_{5}} &=~  - { \nfr {1}{3}} \,{b_{5}} +
{ \nfr {1}{3}} \,{b_{2}} - { \nfr {1
}{3}} \,{b_{3}} - { \nfr {2}{3}} \,{b_{6}} \qquad
{v\cdot\lambda_{6}} ~=~ { \nfr {1}{3}} \,{b_{5}} +
{ \nfr {2}{3}} \,{b_{4}} + { \nfr {1
}{3}} \,{b_{1}} + { \nfr {2}{3}} \,{b_{2}} \cr
{v\cdot\lambda_{7}} &=~ { \nfr {1}{3}} \,{b_{5}} -
{ \nfr {1}{3}} \,{b_{4}} + { \nfr {1
}{3}} \,{b_{1}} + { \nfr {2}{3}} \,{b_{2}} \qquad~~\,
{v\cdot\lambda_{8}} ~=~  - { \nfr {2}{3}} \,{b_{4}} -
{ \nfr {1}{3}} \,{b_{1}} + { \nfr {1
}{3}} \,{b_{3}} + { \nfr {2}{3}} \,{b_{6}} \cr
{v\cdot\lambda_{9}} &=~ { \nfr {1}{3}} \,{b_{4}} -
{ \nfr {1}{3}} \,{b_{1}} - { \nfr {1
}{3}} \,{b_{6}} + { \nfr {1}{3}} \,{b_{3}} \qquad~\,
{v\cdot\lambda_{10}} ~=~  - { \nfr {2}{3}} \,{b_{4}} -
{ \nfr {1}{3}} \,{b_{1}} - { \nfr {1
}{3}} \,{b_{6}} + { \nfr {1}{3}} \,{b_{3}} \cr
{v\cdot\lambda_{11}} &=~ { \nfr {1}{3}} \,{b_{4}} -
{ \nfr {1}{3}} \,{b_{1}} - { \nfr {2
}{3}} \,{b_{3}} - { \nfr {1}{3}} \,{b_{6}} \qquad~\,
{v\cdot\lambda_{12}} ~=~  - { \nfr {2}{3}} \,{b_{4}} -
{ \nfr {1}{3}} \,{b_{1}} - { \nfr {2
}{3}} \,{b_{3}} - { \nfr {1}{3}} \,{b_{6}} \cr
{v\cdot\lambda_{13}} &=~ { \nfr {2}{3}} \,{b_{5}} +
{ \nfr {1}{3}} \,{b_{2}} + { \nfr {1
}{3}} \,{b_{6}} + { \nfr {2}{3}} \,{b_{3}} \qquad~\,
{v\cdot\lambda_{14}} ~=~ { \nfr {2}{3}} \,{b_{5}} +
{ \nfr {1}{3}} \,{b_{2}} - { \nfr {1
}{3}} \,{b_{3}} + { \nfr {1}{3}} \,{b_{6}} \cr
{v\cdot\lambda_{15}} &=~  - { \nfr {1}{3}} \,{b_{5}} +
{ \nfr {1}{3}} \,{b_{2}} + { \nfr {1
}{3}} \,{b_{6}} + { \nfr {2}{3}} \,{b_{3}} ~~~~~
{v\cdot\lambda_{16}} ~=~  - { \nfr {1}{3}} \,{b_{5}} +
{ \nfr {1}{3}} \,{b_{2}} - { \nfr {1
}{3}} \,{b_{3}} + { \nfr {1}{3}} \,{b_{6}} \cr
{v\cdot\lambda_{17}} &=~ { \nfr {2}{3}} \,{b_{5}} +
{ \nfr {1}{3}} \,{b_{2}} - { \nfr {1
}{3}} \,{b_{3}} - { \nfr {2}{3}} \,{b_{6}} \qquad~\,
{v\cdot\lambda_{18}} ~=~  - { \nfr {1}{3}} \,{b_{5}} -
{ \nfr {2}{3}} \,{b_{2}} + { \nfr {2
}{3}} \,{b_{3}} + { \nfr {1}{3}} \,{b_{6}} \cr
{v\cdot\lambda_{19}} &=~  - { \nfr {1}{3}} \,{b_{5}} -
{ \nfr {2}{3}} \,{b_{2}} - { \nfr {1
}{3}} \,{b_{3}} + { \nfr {1}{3}} \,{b_{6}} ~~~~~
{v\cdot\lambda_{20}} ~=~  - { \nfr {1}{3}} \,{b_{5}} -
{ \nfr {2}{3}} \,{b_{2}} - { \nfr {2
}{3}} \,{b_{6}} - { \nfr {1}{3}} \,{b_{3}} \cr
{v\cdot\lambda_{21}} &=~ { \nfr {1}{3}} \,{b_{5}} +
{ \nfr {2}{3}} \,{b_{4}} + { \nfr {1
}{3}} \,{b_{1}} - { \nfr {1}{3}} \,{b_{2}} \qquad~\,
{v\cdot\lambda_{22}} ~=~ { \nfr {1}{3}} \,{b_{5}} -
{ \nfr {1}{3}} \,{b_{4}} + { \nfr {1
}{3}} \,{b_{1}} - { \nfr {1}{3}} \,{b_{2}} \cr
{v\cdot\lambda_{23}} &=~  - { \nfr {2}{3}} \,{b_{5}} +
{ \nfr {2}{3}} \,{b_{4}} + { \nfr {1
}{3}} \,{b_{1}} - { \nfr {1}{3}} \,{b_{2}} ~~~~~
{v\cdot\lambda_{24}} ~=~  - { \nfr {2}{3}} \,{b_{5}} -
{ \nfr {1}{3}} \,{b_{4}} + { \nfr {1
}{3}} \,{b_{1}} - { \nfr {1}{3}} \,{b_{2}} \cr
{v\cdot\lambda_{25}} &=~ { \nfr {1}{3}} \,{b_{4}} +
{ \nfr {2}{3}} \,{b_{1}} + { \nfr {1}{3}} \,{b_{3}} + { \nfr {2}{3}}
\,{b_{6}} \qquad~\, {v\cdot\lambda_{26}} ~=~ { \nfr {1}{3}} \,{b_{4}} +
{ \nfr {2}{3}} \,{b_{1}} + { \nfr {1
}{3}} \,{b_{3}} - { \nfr {1}{3}} \,{b_{6}} \cr
{v\cdot\lambda_{27}} &=~ { \nfr {1}{3}} \,{b_{4}} +
{ \nfr {2}{3}} \,{b_{1}} - { \nfr {2
}{3}} \,{b_{3}} - { \nfr {1}{3}} \,{b_{6}}}
}
This gives the matrix $v\cdot H$ in terms of the
Fayet-Iliopoulos terms and thus the Casimir operators \Ps.
In our normalization, the weights have
length squared $\lambda\cdot\lambda = \frac23$, which corresponds to
$\Tr\left(T_a T_b \right) = 3 \delta_{ab}$ in the
fundamental representation.
\listrefs

\end